\documentclass[10pt,aps,prb,reprint,superscriptaddress,floatfix]{revtex4-1}

\usepackage[UKenglish]{babel}
\usepackage[utf8]{inputenc}
\usepackage[T1]{fontenc}

\usepackage{graphicx}
\usepackage{amssymb}
\usepackage{amsmath}
\usepackage{hyperref}

\hypersetup{pdfauthor={C. Hubig, I. P. McCulloch, U. Schollwoeck,
    F. A. Wolf},pdftitle={A Strictly Single-Site DMRG Algorithm}}

\usepackage{enumitem}
\usepackage{color}
\usepackage[amssymb]{SIunits}
\providecommand{\e}[1]{\ensuremath{\times 10^{#1}}}

\begin{document}
\title{A Strictly Single-Site DMRG Algorithm with Subspace Expansion}
\author{C.\ Hubig}
\email{c.hubig@physik.uni-muenchen.de}
\affiliation{Department of Physics and Arnold Sommerfeld Center for
  Theoretical Physics, Ludwig-Maximilians-Universit\"at M\"unchen,
  Theresienstrasse 37, 80333 M\"unchen, Germany}
\author{I.\ P.\ McCulloch}
\affiliation{Centre for Engineered Quantum Systems, School of Physical
  Sciences, The University of Queensland, Brisbane, Queensland 4072,
  Australia}
\author{U.\ Schollw\"ock}
\affiliation{Department of Physics and Arnold Sommerfeld Center for
  Theoretical Physics, Ludwig-Maximilians-Universit\"at M\"unchen,
  Theresienstrasse 37, 80333 M\"unchen, Germany}
\author{F.\ A.\ Wolf}
\affiliation{Department of Physics and Arnold Sommerfeld Center for
  Theoretical Physics, Ludwig-Maximilians-Universit\"at M\"unchen,
  Theresienstrasse 37, 80333 M\"unchen, Germany}
\begin{abstract}
  We introduce a strictly single-site DMRG algorithm based on the
  subspace expansion of the Alternating Minimal Energy (AMEn)
  method. The proposed new MPS basis enrichment method is sufficient
  to avoid local minima during the optimisation, similarly to the
  density matrix perturbation method, but computationally
  cheaper. Each application of $\hat H$ to $|\Psi\rangle$ in the
  central eigensolver is reduced in cost for a speed-up of $\approx (d
  + 1)/2$, with $d$ the physical site dimension. Further speed-ups
  result from cheaper auxiliary calculations and an often greatly
  improved convergence behaviour. Runtime to convergence improves by
  up to a factor of 2.5 on the Fermi-Hubbard model compared to the
  previous single-site method and by up to a factor of 3.9 compared to
  two-site DMRG. The method is compatible with real-space
  parallelisation and non-abelian symmetries.
\end{abstract}
\date{\today}
\maketitle
\section{\label{sec:intro}Introduction}

Since its introduction in 1993,\cite{white92:_densit, white93:_densit}
the Density Matrix Renormalisation Group method (DMRG) has seen
tremendous use in the study of one-dimensional
systems.\cite{schollwoeck05, schollwoeck11} Various improvements such
as real-space parallelisation,\cite{stoudenmire13:_real} the use of
abelian and non-abelian symmetries\cite{mcculloch02:_abelian} and
multi-grid methods\cite{dolfi12:_multig_algor_tensor_networ_states}
have been proposed. Most markedly, the
introduction\cite{white05:_densit} of density matrix perturbation
steps allowed the switch from two-site DMRG to single-site DMRG in
2005, which provided a major speed-up and improved convergence in
particular for systems with long-range interactions.

Nevertheless, despite some
progress,\cite{yan11:_spin_liquid_groun_state_s,
  depenbrock12:_natur_spin_liquid_groun_state,
  stoudenmire12:_study_two_dimen_system_densit} (nearly)
two-dimensional systems, such as long cylinders, are still a hard
problem for DMRG.  The main reason for this is the different scaling
of entanglement due to the area
law:\cite{vidal03:_entan_quant_critic_phenom, eisert10:_colloq} In one
dimension, entanglement and hence matrix dimensions in DMRG are
essentially size-independent for ground states of gapped systems,
whereas in two dimensions, entanglement grows linearly and matrix
dimensions roughly exponentially with system width.

As a result, the part of the Hilbert space considered by DMRG during
its ground state search increases dramatically, resulting mainly in
three problems: firstly, the DMRG algorithm becomes numerically more
challenging as the sizes of matrices involved grow (we will assume
matrix-matrix multiplications to scale as $O(m^3)$ throughout the
paper). Secondly, the increased search space size makes it more likely
to get stuck in local minima. Thirdly, while sequential updates work
well in 1-D chains with short-range interactions, nearest-neighbour
sites in the 2-D lattice can be separated much farther in the DMRG
chain. Therefore, improvements to the core DMRG algorithm are still
highly worthwhile.

In this paper, we will adopt parts of the AMEn
method\cite{dolgov14:_alter_minim_energ_method_linear} developed in
the tensor train/numerical linear algebra community to construct a
strictly single-site DMRG algorithm that works without accessing the
(full) reduced density matrix. Compared to the existing
\emph{centermatrix wavefunction formalism}
(CWF),\cite{mcculloch07:_from} we achieve a speed-up of $\approx
(d+1)/2$ during each application of $\hat H$ to $|\Psi\rangle$ in the
eigensolver during the central optimisation routine, where $d$ is the
dimension of the physical state space on each site.

The layout of this paper is as follows: Section \ref{sec:notation}
will establish the notation. Section \ref{sec:cwf} will recapitulate
the density matrix perturbation method and the CWF. Section
\ref{sec:subexp} will introduce the subspace expansion method and the
heuristic expansion term with a simple two-spin example.  The
\underline{s}trictly \underline{s}ingle-\underline{s}ite DMRG
algorithm (DMRG3S) will be presented in Section \ref{sec:dmrg3s}
alongside a comparison with the existing CWF.  As both the original
perturbation method and the heuristic subspace expansion require a
\emph{mixing factor} $\alpha$,\cite{white05:_densit} Section
\ref{sec:alpha} describes how to adaptively choose $\alpha$ for
fastest convergence. Numerical comparisons and examples will be given
in Section \ref{sec:numexps}.

\section{\label{sec:notation}DMRG Basics}

The notation established here closely follows the review article Ref.\
\onlinecite{schollwoeck11}. Consider a state $|\Psi\rangle$ of a
system of $l$ sites. Each site has a physical state dimension $d_i$,
e.g.\ $\forall i: d_i = 3$, $l = 50$ for a system of $50$ $S=1$ spins:
\begin{equation}
  |\Psi\rangle = \sum_{\sigma_1 \ldots \sigma_l} c_{\sigma_1 \ldots
  \sigma_l} |\sigma_1 \ldots \sigma_l\rangle \quad.
\end{equation}
In practice, the dimension of the physical basis is usually constant,
$\forall i: d_i = d$, but we will keep the subscript to refer to one
specific basis on site $i$ where necessary.

It is then possible to decompose the coefficients $c_{\sigma_1,
  \ldots, \sigma_l}$ as a series of rank-3 tensors $M_1, \ldots, M_l$
of size $(d_i,m_{i-1},m_i)$ respectively, with $m_0 = m_l = 1$. The
coefficient $c_{\sigma_1, \ldots, \sigma_l}$ can then be written as
the matrix product of the corresponding matrices in $M_1, \ldots,
M_l$:
\begin{equation}
  |\Psi\rangle = \sum_{\sigma_1 \ldots \sigma_l}
  \underbrace{M_1^{\sigma_1} \cdots M_l^{\sigma_l}}_{c_{\sigma_1
  \ldots \sigma_l}} |\sigma_1 \ldots \sigma_l\rangle
  \label{eq:psi-as-mps-general}\quad.
\end{equation}
The maximal dimension $m = \mathrm{max}_{i}\left\{m_i\right\}$ is
called the \emph{MPS bond dimension}. In typical one-dimensional
calculations, $m = 200$, but for e.g.\ $32 \times 5$ cylinders, $m >
5000$ is often necessary. It is in these numerically demanding cases
that our improvements are of particular relevance.

Similarly, a Hamiltonian operator can be written as a \emph{matrix
  product operator} (MPO), where each tensor $W_i$ is now of rank 4,
namely $(d_i,d_i,w_{i-1},w_i)$:
\begin{equation}
  \hat H = \sum_{\substack{\sigma_1\ldots\sigma_l\\\tau_1
  \ldots\tau_l}} W_1^{\sigma_1\tau_1} \cdots W_l^{\sigma_l\tau_l}
  |\sigma_1\ldots\sigma_l\rangle\langle\tau_1\ldots\tau_l|.
\end{equation}
$w = \mathrm{max}_i\left\{w_i\right\}$ is called the \emph{MPO bond
  dimension}. We will usually assume that for most $i$, $m_i = m$ and
$w_i = w$. In practice, this holds nearly everywhere except at the
ends of the chain, where the $m_i$ grow exponentially from $1$ to
$m$. The basis of $M_i$ ($W_i$) of dimension $m_{i-1}$ ($w_{i-1}$) is
called the left-hand side (LHS) basis, whereas the basis of dimension
$m_i$ ($w_i$) is the right-hand side (RHS) basis of this tensor. For
simplicity, $m_i$, $d_i$ and $w_i$ can also refer to the specific
basis (and not only its dimension) when unambiguous.

Instead of $M_i$, we will also write $A_i$ ($B_i)$ for a left (right)
normalised MPS tensor:
\begin{align}
  \sum_{\sigma_i} A_i^{\sigma_i\dagger} A_i^{\sigma_i} & = \mathbb{I} \\
  \sum_{\sigma_i} B_i^{\sigma_i} B_i^{\sigma_i\dagger} & = \mathbb{I} \quad.
\end{align}
If we then define the contractions
\begin{align}
  l_i & = \left(A^{\sigma_1}_1 \cdots A^{\sigma_{i-1}}_{i-1}
    M^{\sigma_i}_i\right) \in (d_1,\ldots,d_i,m_i) \\
  r_i & = \left(M^{\sigma_i}_i B^{\sigma_{i+1}}_{i+1} \cdots
    B^{\sigma_l}_l\right) \in (m_{i-1},d_i,\ldots,d_{l}) \quad,
\end{align}
we can rewrite $|\Psi\rangle$ from \eqref{eq:psi-as-mps-general} as
\begin{equation}
  |\Psi\rangle = \sum_{\sigma_1\ldots\sigma_l} l_i r_{i+1} |\sigma_1
  \ldots \sigma_i\rangle \otimes |\sigma_{i+1}\ldots\sigma_l\rangle
  \label{eq:psi-decomp-bond} \quad.
\end{equation}
That is, when only considering one specific bond $(i,i+1)$, the left
and right MPS bases at this bond are built up from the states
generated by the MPS tensor chains to the left and right of the
bond. Individual elements of an MPS basis are therefore called
``state''.

Furthermore, define $L_0 = 1$ and $L_i = L_{i-1} A_i^\dagger W_i A_i$
with summation over all possible indices. Similarly, $R_{l+1} = 1$ and
$R_i = R_{i+1} B_i^\dagger W_i B_i$. With these contractions, it is
possible to write
\begin{equation}
  \langle \Psi | \hat H | \Psi \rangle = L_{i-1} M_i^\dagger W_i M_i R_{i+1}
\end{equation}
for any $i \in \left[0,l\right]$.

DMRG then works by \emph{sweeping} over the system multiple times.
During each sweep, each site tensor $M_i$ is sequentially
\emph{updated} once with each update consisting of one optimisation
step via e.g.\ a sparse eigensolver and possibly one \emph{enrichment}
step during which the left or right MPS basis of $M_i$ is changed in
some way. Depending on the exact implementation, updates may work on
one (single-site DMRG) or two sites (two-site DMRG) at a time. The
enrichment step may be missing or implemented via Density Matrix
Perturbation or Subspace Expansion.

\section{\label{sec:cwf}Perturbation Step and
Centermatrix Wavefunction Formalism (CWF)}

\subsection{\label{sec:cwf:conv}Convergence Problems of Single-Site DMRG}
During single-site DMRG, only a single MPS tensor $M_i$ on site $i$ is
optimised at once. Compared to two-site DMRG, the search space is
reduced by a factor of $d \approx 2 \ldots 5$, leading to a speed-up of
at least $O(d)$ per iteration.\cite{white05:_densit} However, since
the left and right bases of the tensors $M_i$ are fixed and defined by
the environment ($l_{i-1}$ and $r_{i+1}$), this approach is likely to
get stuck. While also occurring if there are no symmetries implemented
on the level of the MPS, this issue is most easily visible if one
considers $U(1)$ symmetries:\cite{schollwoeck11} assume that all basis
states to the right of the RHS bond of $M_i$ transform as some quantum
number $s_z$. If we now target a specific sector, e.g.\ $S_z = 0$
overall, then on the LHS of this bond (i.e.\ from the left edge up to
and including $M_i$), all states must transform as $-s_z$.  In this
configuration, it is impossible for a local change of $M_i$ to add a
new state that transforms as, say, $s^\prime_z$, to its right basis
states, as there would be no corresponding state $-s^\prime_z$ to the
right of that bond, rendering the addition of the state moot from the
perspective of the local optimiser, as its norm will be zero
identically. A concrete example of this issue is given in Section
\ref{sec:numexps:stuckspins}.

DMRG is a variational approach on the state space available to MPS of
a given bond dimension. As such, the algorithm must converge into
either the global or a local minimum of the energy in this state
space. Hence, we will call all cases where DMRG converges on an energy
substantially higher than the minimal energy achievable with the
allowed MPS bond dimension cases where DMRG is stuck in \emph{local
  minima}.

\subsection{\label{sec:cwf:rhopert}Density Matrix Perturbation}

This convergence problem has been solved by White
(2005).\cite{white05:_densit} In the following, we will assume a
left-to-right sweep, sweeping in the other direction works similarly,
but on the left rather than right bonds.  After the local optimisation
of the tensor $M_i$, the reduced density matrix
\begin{equation}
  \rho_{i,R} = l_{i-1} M_i M_i^\dagger l_{i-1}^\dagger
\end{equation}
is built on the next bond. This is the reduced density matrix
resulting from tracing out the part of the system to the left of bond
$(i,i+1)$.

$\rho_{i,R}$ is then perturbed as
\begin{equation}
  \rho_{i,R} \to \rho_{i,R}' = \rho_{i,R} + \alpha \mathrm{Tr}
  \left( L_i \rho_{i,R} L_i^\dagger\right) \quad. \label{eq:densitypert}
\end{equation}
The new $\rho_{i,R}'$ is then used to decide on a new set of basis
states on the RHS of $M_i$, with the inverse mapping from the new to
the old basis being multiplied into each component of $B_{i+1}$. The
mixing factor $\alpha$ is a small scalar used to control the
perturbation. A new scheme to find the optimal choice of $\alpha$ is
discussed in Section \ref{sec:alpha}.

\subsection{Centermatrix Wavefunction Formalism (CWF)}

In a standard single-site DMRG calculation, the reduced density matrix
$\rho_{i,R}$ is never used. More importantly, even building
$\rho_{i,R}$ on a given bond $(i,i+1)$ will not yield a density matrix
that can be used in \eqref{eq:densitypert}, as it only contains the
$m_i$ states existing on that bond already without knowledge of the
$m_{i-1}$ states on the bond one step to the left. In other words, it
is not possible to choose the optimal set $\widetilde{m_i}$ based only
on $m_i$, rather, one requires also $d_i$ and $m_{i-1}$.

The centermatrix wavefunction formalism\cite{mcculloch07:_from} was
developed to cope with this problem. Given a site tensor $M_i \in
(d_i,m_{i-1},m_i)$ on a left-to-right sweep, it introduces a
``centermatrix'' $C_{i,R} \in (d_im_{i-1},m_i)$ and replaces the
original site tensor as
\begin{equation}
  M_i \to A_i \in (d_i,m_{i-1},d_im_{i-1}) \textrm{ s.t.\ } M_i = A_i C_{i,R}.
\end{equation}
$A_i$ is constructed to be left-orthogonal and is essentially an
identity matrix mapping the left basis $m_{i-1}$ and the physical
basis $d_i$ onto a complete basis containing all $d_im_{i-1}$ states
on its right. The new basis is ``complete'' in the sense that all
states reachable from the left bond basis $m_{i-1}$ and the local
physical basis $d_i$ are contained within it.

The contents of $M_i$ are placed in $C_{i,R}$ accordingly and the
original state remains unchanged. The reduced density matrix is then
$\rho_{i,R} = C_{i,R} C_{i,R}^\dagger$ and has access to all
$d_im_{i-1}$ states, as required above. A perturbation of $\rho_{i,R}$
according to \eqref{eq:densitypert} hence allows the introduction of
new states.

The DMRG optimisation step can work on $C_{i,R}$ alone, with $L_i$
built prior to optimisation of $C_{i,R}$ from the expanded
$A_i$. During each eigensolver step, the effective Hamiltonian on site
$i$ has to be applied onto $C_{i,R}$. The application is done by
contraction of $L_i \in (w,d_im_{i-1},d_im_{i-1})$, $R_{i+1} \in
(w,m_i,m_i)$ and $C_{i,R} \in (d_im_{i-1},m_i)$ at cost $O(w (d^2 + d)
m^3)$ per step. After optimisation, the perturbation is added. Its
computational cost is dominated by the calculation of $\alpha
\mathrm{Tr}\{L_i \rho_{i,R} L_i^\dagger\}$ at $O(w d^3 m^3)$. The bond
between $A_i$ and $C_{i,R}$ can then be truncated down to $m$ using
$\rho'_{i,R}$ and the remaining parts of $C_{i,R}$ are multiplied into
$B_{i+1}$ to the right.

The resulting algorithm converges quickly for one-dimensional problems
and performs reasonably well for small cylinders. However, both the
cost of the applications of $\hat H$ to $|\Psi\rangle$ as $O(w (d^2+d)
m^3)$ as well as the large density matrix $\rho \in (dm,dm)$ cause
problems if $m$ and $w$ become large.

\section{\label{sec:subexp}Subspace Expansion}

The idea of using \emph{subspace expansion} instead of density matrix
perturbation originates
\cite{dolgov14:_alter_minim_energ_method_linear, dolgov13:_one} in the
tensor train/numerical linear algebra community. There, a stringent
proof was given regarding the convergence properties of this method
when the local tensor $Z_i$ of the residual 
\begin{equation}
  | Z \rangle \equiv \hat H | \Psi \rangle - E | \Psi \rangle =
  \sum_{\sigma_1\ldots\sigma_l} Z_1^{\sigma_1}
  \cdots Z_l^{\sigma_l}|\sigma_1 \ldots \sigma_l\rangle
\end{equation}
is used as the expansion term. Here, we will only use the method of
subspace expansion and substitute a numerically much more cheaply
available expansion term.

The following section is divided into three parts: firstly, we will
explain the concept of subspace expansion acting on two neighbouring
MPS tensors $M_i$, $M_{i+1}$. Secondly, the expansion term employed in
DMRG3S is introduced and motivated. Thirdly, a simple example is
described.

\subsection{\label{sec:subexp-subexp}Subspace Expansion with an
  Arbitrary Expansion Term}

In the following, we will describe subspace expansion of the RHS basis
of the current working tensor, as it would occur during a
left-to-right sweep.

Assume a state $|\Psi\rangle$ described by a set of tensors $\left\{
  A_1 , \ldots , A_{i-1}, M_i , B_{i+1} , \ldots, B_l \right\}$. At
the bond $(i,i+1)$, we can then decompose the state as a sum over left
and right basis states as in Eq.\ \eqref{eq:psi-decomp-bond}.

Now we \emph{expand} the tensor $M_i \in (d,m_{i-1},m_i)$ by some
expansion term $P_i \in (d,m_{i-1},m_{P_i})$ for each individual
physical index component:
\begin{equation}
  M_i^{\sigma_i} \to \tilde M_i^{\sigma_i} = \left[ \begin{matrix}
  M_i^{\sigma_i} & P_i^{\sigma_i} \end{matrix} \right] \quad.
\end{equation}
This effectively expands the RHS MPS basis of $M_i$ from $m_i$ to $m_i
+ m_{P_i}$. Similarly, expand the components of $B_{i+1} \in
(d,m_i,m_{i+1})$ with zeros:
\begin{equation}
  B_{i+1}^{\sigma_{i+1}} \to \tilde B_{i+1}^{\sigma_{i+1}} = \left[
  \begin{matrix} B_{i+1}^{\sigma_{i+1}} \\ 0 \end{matrix}\right] \quad.
\end{equation}
The appropriately-sized block of zeros only multiplies with the
expansion term $P_i^{\sigma_i}$. In terms of a decomposition as in
\eqref{eq:psi-decomp-bond}, this is equivalent to
\begin{equation}
  |\Psi\rangle = \sum_{\sigma_1,\ldots,\sigma_l} \left[ l_i \; p \right]
  \left[ \begin{matrix} r_{i+1} \\ 0 \end{matrix} \right] |\sigma_1 \ldots
  \sigma_i\rangle \otimes |\sigma_{i+1},\ldots,\sigma_l\rangle
  \label{eq:subexp-post}
\end{equation}
where $p$ is the result of multiplying $l_{i-1}$ and $P_i$, with the
$0$ in the second expression similarly resulting from the $0$ in
$B_{i+1}$. While the state $|\Psi\rangle$ remains unchanged, the local
optimiser on the new site $B_{i+1}$ can now choose the initially-zero
components differently if so required: The necessary flexibility in
the left-/right basis states to escape local minima has been achieved
without referring to the density matrix.

Note that while orthonormality of $B_{i+1}$ is lost, we do not need it
between the enrichment step on site $i$ and the optimisation step on
site $i+1$. The orthonormality of $M_i$ can be restored via singular
value decomposition as usual. Furthermore, it is usually necessary to
truncate the RHS basis of $\tilde M_i$ down from $m_i+m_{P_i}$ to $m$
immediately following the expansion: this preserves the most
relevant states of the expansion term while avoiding an exponential
explosion of bond dimensions.

When sweeping from right to left, the left rather than right MPS basis
of the current working tensor is expanded, with the left tensor
$A_{i-1}$ being zero-padded as opposed to the right tensor $B_{i+1}$:
\begin{align}
  M_i^{\sigma_i} & \to \tilde M_i^{\sigma_i} \; \; = \left[
    \begin{matrix} M_i^{\sigma_i} \\ P_i^{\sigma_i} \end{matrix} \right] \\
  A_{i-1}^{\sigma_{i-1}} & \to \tilde A_{i-1}^{\sigma_{i-1}} = \left[
    \begin{matrix} A_{i-1}^{\sigma_{i-1}} & 0 \end{matrix}\right] \quad.
\end{align}

\subsection{\label{sec:subexp-heuristic}Expansion Term}
Using the exact residual as the expansion term is computationally
expensive: The term $\hat H |\Psi\rangle$ can be updated locally and
is mostly unproblematic, but the subtraction of $E |\Psi\rangle$ and
subsequent re-orthonormalisation is costly and has to be done after
each local optimisation, as the current value of $E$ changes. This
exact calculation is hence only possible for $m \approx 100$, which is
far too small to tackle difficult two-dimensional problems.

Instead, we propose the very cheaply available terms
\begin{equation}
  P_i = \alpha L_{i-1} M_i W_i \in (d_i,m_{i-1},w_i m_i) \label{eq:subexp-ourt}
\end{equation}
to be used during left-to-right sweeps and $P_i = \alpha R_{i+1} M_i
W_i$ for use during right-to-left sweeps with some scalar mixing
factor $\alpha$. In the regime where the exact residual can be
computed, these terms work essentially equally well.

This expression for $P_i$ can be heuristically motivated as follows:
\eqref{eq:subexp-ourt} is equivalent to the partial projection of $H
|\Psi\rangle$ onto $|\Psi\rangle$ to the left of the current
bond. Hence, in the ground state and ignoring numerical errors, the
RHS basis of this $P_i$ is identical to that of $M_i$.  Truncation
from $m_i + m_{P_i}$ to $m_i$ is then possible without inducing
errors.

Numerically, it seems possible to choose $\alpha$ arbitrarily large
without hindering convergence or perturbing the state too much in
simple (one-dimensional) problems. However, if the chosen maximal bond
dimension $m$ is insufficient to faithfully capture the ground state
of the given system, $\alpha$ has to be taken to zero eventually to
allow convergence. Otherwise, $P_i$ will continuously add new states
and disturb the result of the eigensolver, which is optimal at this
specific value of $m$ but not an eigenstate of $\hat H$ yet.

The cost of a single subspace expansion is $O(w d m^3 + w^2 d^2 m^2)$
for the calculation of $P_i$, potentially $O(2dwm^2)$ for the addition
to $M_i$ and $B_{i+1}$ respectively and $O(dw^2m^3 + d^2m^2)$ for the
SVD of an $(dm,wm)$ matrix formed from $\tilde M_i$. If we restrict
the SVD to $m$ singular values, then the resulting matrices will be of
dimension $(dm,m)$, $(m,m)$ and $(m,wm)$ respectively. The first can
be reformed into $\tilde A_i$ at cost $O(dm^2)$ and the second and
third multiplied into $B_{i+1}$ at cost $O(m^3dw + m^3d)$. The total
cost of this step is dominated by the cost of the SVD at $O(dw^2m^3)$,
which is still cheaper than the calculation of the perturbation term
in \eqref{eq:densitypert}, not considering the other costs associated
to using the density matrix for truncation.

\subsection{\label{seq:subexp-2x2}Subspace Expansion at the Example
  of a $d = l = 2$ Spin System}
In the following, we will demonstrate and illustrate the method of
subspace expansion at the simple example of a system of two spins with
$S = \frac{1}{2}$ from $m = 1$ to $m = 2$ as it would occur during a
left-to-right sweep.

Assume the Hamiltonian
\begin{align}
  H & = S^1_x S^2_x + S^1_y S^2_y + S^1_z S^2_z \\ 
    & = \frac{1}{2} \left\{ S^1_+S^2_- + S^1_-S^2_+ \right\} + S^1_zS^2_z \quad
\end{align}
with MPO-components
\begin{align}
  W_1 & = \left[ \frac{1}{\sqrt{2}} S_+ \quad
    \frac{1}{\sqrt{2}} S_- \quad S_z \right] \\
  W_2 & = \left[ \frac{1}{\sqrt{2}} S_- \quad
    \frac{1}{\sqrt{2}} S_+ \quad S_z \right]^T \quad.
\end{align}
Let the initial state be an $m=1$ MPS, described by components
\begin{align}
  A_1^\uparrow = [ a ] \quad & A_1^\downarrow = \left[ \sqrt{1-a^2} \right] \\
  B_2^\uparrow = [ b ] \quad & B_2^\downarrow = \left[ \sqrt{1-b^2} \right]
\end{align}
where square brackets denote matrices in the MPS bond indices. Due to
the standard normalisation constraints, there are only two free scalar
variables here, $a$ and $b$.

Subspace expansion of $A_1$ is straightforward (keep in mind that $L_0
\equiv 1$ for convenience):
\begin{align}
  P_1^{\tau_1} & = \sum_{\sigma_1} W_1^{\tau_1 \sigma_1} A_1^{\sigma_1} \\
  P_1^\uparrow & = W_1^{\uparrow\uparrow} A_1^\uparrow +
  W_1^{\uparrow\downarrow} A_1^\downarrow \\ & = \left[ \begin{matrix}
    \frac{\sqrt{1-a^2}}{\sqrt{2}} & 0 & a \end{matrix} \right] \\
  P_1^\downarrow & = W_1^{\downarrow\uparrow} A_1^\uparrow +
  W_1^{\downarrow\downarrow} A_1^\downarrow \\ & = \left[ \begin{matrix}
    0 & \frac{a}{\sqrt{2}} & - \sqrt{1-a^2} \end{matrix} \right]
\end{align}
resulting in $A^\prime_1$ and $B^\prime_2$ directly after the expansion:
\begin{align}
  A_1^{\prime\uparrow}   & = \left[ \begin{matrix} a &
    \frac{\sqrt{1-a^2}}{\sqrt{2}} & 0 & a \end{matrix} \right] \\
  A_1^{\prime\downarrow} & = \left[ \begin{matrix} \sqrt{1-a^2} & 0 &
    \frac{a}{\sqrt{2}} & - \sqrt{1-a^2} \end{matrix} \right] \\
  B_2^{\prime\uparrow}   & = \left[ \begin{matrix} b \\ 0 \\ 0 \\ 0 \end{matrix}
    \right] \quad B_2^{\prime\downarrow} = \left[
    \begin{matrix} \sqrt{1-b^2} \\ 0 \\ 0 \\ 0 \end{matrix} \right] \quad.
\end{align}
Normalising $A^\prime_1$ via a singular value decomposition as
$A^\prime_1 \to A^{\prime\prime}_1 SV^\dagger$ and multiplying
$SV^\dagger B^{\prime}_2 \to B^{\prime\prime}_2$ gives:
\begin{align}
  A_1^{\prime\prime\uparrow}
    & = \left[ \begin{matrix} 1 & 0 \end{matrix} \right] \\
  A_1^{\prime\prime\downarrow}
    & = \left[ \begin{matrix} 0 & 1 \end{matrix} \right] \\
  SV^\dagger & = \left[ \begin{matrix}
    a & \frac{\sqrt{1-a^2}}{\sqrt{2}} & 0 & a \\
    \sqrt{1-a^2} & 0 & \frac{a}{\sqrt{2}} & - \sqrt{1-a^2}
  \end{matrix}\right] \\
  B_2^{\prime\prime\uparrow} & = \left[
    \begin{matrix} a b \\ \sqrt{1-a^2} b \end{matrix} \right] \\
  B_2^{\prime\prime\downarrow} & = \left[
    \begin{matrix} a \sqrt{1-b^2} \\ \sqrt{1-a^2} \sqrt{1-b^2}
    \end{matrix}\right]\quad.
\end{align}

As expected, the final state $|\Psi\rangle = \sum_{\sigma_1 \sigma_2}
A_1^{\prime\prime\sigma_1} B_2^{\prime\prime\sigma_2}$ is still entirely
unchanged, but there is now a one-to-one correspondence between the
four entries of $B^{\prime\prime}_2$ and the coefficients
$c_{\left\{\uparrow,\downarrow\right\},\left\{\uparrow,\downarrow\right\}}$
in the computational basis, making the optimisation towards $c_{ii} =
0, c_{i \neq j} = \frac{1}{\sqrt{2}}$ trivial.

\section{\label{sec:dmrg3s}Strictly Single-Site DMRG}

We can now combine standard single-site DMRG (e.g.\
Ref. \onlinecite{schollwoeck11}, p.\ 67) with the subspace expansion
method as a way to enrich the local state space, leading to a
\underline{s}trictly \underline{s}ingle-\underline{s}ite DMRG
implementation (DMRG3S) that works without referring to the density
matrix at any point.

With the notation from Section \ref{sec:notation}, the steps follow
mostly standard single-site DMRG. In an outermost loop, the algorithm
sweeps over the system from left-to-right and right-to-left until
convergence is reached. Criteria for convergence are e.g.\ diminishing
changes in energy or an overlap close to $1$ between the states at the
ends of subsequent sweeps.

The inner loop sweeps over the system, iterating over and updating the
tensors on each site sequentially. Each local update during a
left-to-right sweep (right-to-left sweeps work analogously) consists
of the following steps:
\begin{enumerate}
\item Optimise the tensor $M_i$: Use an eigensolver targeting the
  smallest eigenvalue to find a solution $(M_i^\star,\lambda^\star)$
  to the eigenvalue problem
  \begin{equation}
    L_{i-1} R_{i+1} W_i M_i = \lambda M_i \quad. \label{eq:es-iter}
  \end{equation}
  $\lambda^\star$ is the new current energy estimate. This first step
  dominates the computational cost.
\item Build $\alpha P_i$ according to \eqref{eq:subexp-ourt} using
  $M_i^\star$. Build an appropriately-sized zero block $0_{i+1}$ after
  the dimensions of $P_i$ are known.
\item Subspace-expand $M_i^\star \to \tilde M_i^\star$ with $\alpha
  P_i$ and $B_{i+1}$ with $0_{i+1}$.
\item Apply a SVD to $\tilde M_i^\star$ and truncate its right basis
  to $m_i$ again, resulting in $\tilde A_i^\star$.
\item Multiply the remainder of the SVD ($SV^\dagger$) into $B_{i+1}
  \to \tilde B_{i+1}$.
\item Build $L_i$ from $\tilde A_i^\star$, $L_{i-1}$ and $W_i$.
\item Calculate a new energy value after truncation based on $L_{i}$,
  $\tilde B_{i+1}$, $W_{i+1}$ and $R_{i+1}$. Use this energy value and
  $\lambda^\star$ to adapt the current value of $\alpha$
  (cf. Section~\ref{sec:alpha}).
\item Continue on site $i+1$.
\end{enumerate}
Of these, step 2 and 3 implement the actual subspace expansion,
whereas all others are identical to standard single-site DMRG.

It is important to note that the only change from standard single-site
DMRG is the addition of an enrichment step via subspace
expansion. Therefore, this method does not interfere with
e.g. real-space parallelised DMRG,\cite{depenbrock13:_tensor,
  stoudenmire13:_real} the use of nonabelian
symmetries\cite{mcculloch02:_abelian, mcculloch07:_from} or multi-grid
methods.\cite{dolfi12:_multig_algor_tensor_networ_states}

To analyse the computational cost, we have to take special care to
ensure optimal ordering of the multiplications during each eigensolver
iteration in \eqref{eq:es-iter}. The problem is to contract $L_{i-1}
R_{i+1} W_i M_i$, with $L_{i-1}$ and $R_{i+1} \in (w,m,m)$, $W_i \in
(d,d,w,w)$ and $M_i \in (d,m,m)$. The optimal ordering is then
$(((L_{i-1}M_i)W_i)R_{i+1})$:
\begin{enumerate}
\item Contract $L_{i-1}$ and $M_{i}$ over the left MPS bond at cost
  $O(mw \cdot m \cdot dm = m^3wd)$.
\item Multiply in $W_i$ over the physical bond of $M_i$ and the left
  MPO bond at cost $O(m^2 \cdot wd \cdot dw = m^2 d^2 w^2)$.
\item Finally contract with $R_{i+1}$ over the right MPO and MPS bonds
  at cost $O(md \cdot wm \cdot m = m^3 dw)$.
\end{enumerate}
The total cost of this procedure to apply $\hat H$ to $|\Psi\rangle$
is $O(2 m^3 wd + d^2 m^2 w^2)$. Assuming large $d^2w/m$ is small, this
gives a speed-up in the eigensolver multiplications of $(d + 1)/2$
over the CWF approach, which takes $O(m^3 w d (d+1))$.

In addition to this speed-up, the subspace expansion is considerably
cheaper than the density matrix perturbation. Since the
perturbation/truncation step can often take up to 30\% of total
computational time, improvements there also have a high impact. At the
same time, the number of sweeps at large $m$ needed to converge does
not seem to increase compared to the CWF approach (cf.\ Section
\ref{sec:numexps}) and sometimes even decreases.

\section{\label{sec:alpha}Adaptive Choice of Mixing Factor}

Both density matrix perturbation and subspace expansion generally
require some small mixing factor $\alpha$ to moderate the
contributions of the perturbation terms. The optimal choice of this
$\alpha$ depends on the number of states available and those required
to represent the ground state, as well as the current speed of
convergence. Too large values for $\alpha$ hinder convergence by
destroying the improvements made by the local optimiser, whereas too
small values lead to the calculation being stuck in local minima with
vital states not added for the reasons given in Section
\ref{sec:cwf:rhopert}. The correct choice of $\alpha$ hence affects
calculations to a large degree, but is also difficult to estimate
before the start of the calculation.

\begin{figure}
  \centering
  \includegraphics[width=\columnwidth]{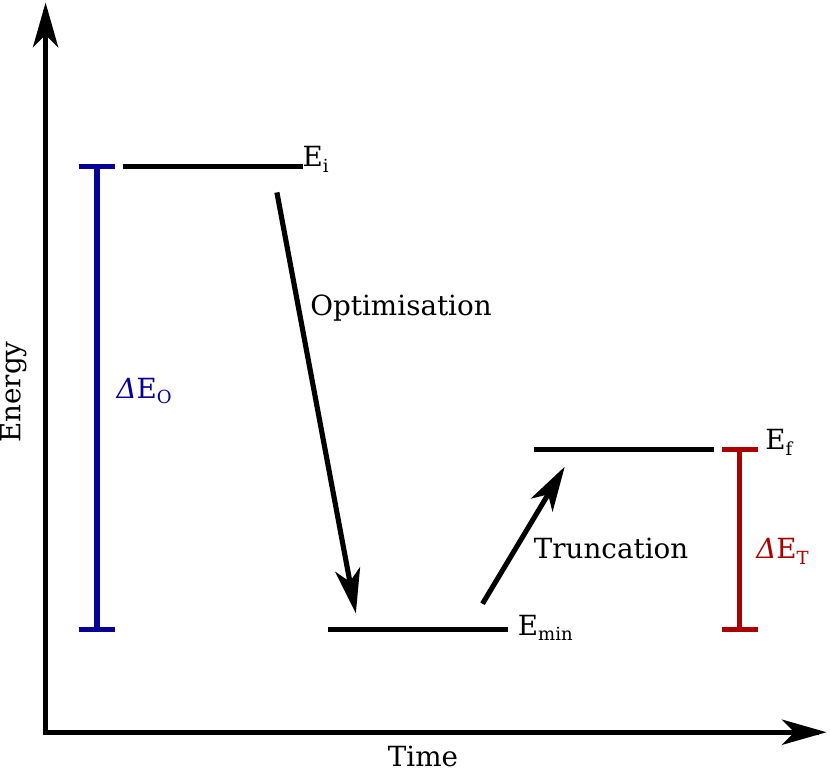}
  \caption{\label{fig:energy-levels-alpha}(Colour online) Energies of
    the state at different points during a single update: Before
    optimisation, the state has some initial energy $E_i$. Local
    optimisation via the eigensolver takes this energy down by $\Delta
    E_O$ to $E_{min}$. Subsequent truncation causes a rise in energy
    by $\Delta E_T$ with the final value at the end of this update
    being $E_f$.}
\end{figure}

Fig.~\ref{fig:energy-levels-alpha} displays the individual steps
within a single update from the energy perspective: Let $\Delta E_O$
denote the gain in energy during the optimisation step and let $\Delta
E_T$ denote the subsequent rise in energy during the truncation
following the enrichment step. $\Delta E_T \neq 0$ only occurs if some
enrichment (either via density matrix perturbation or subspace
expansion) has occurred, otherwise there would be no need for any sort
of truncation. We can hence control the approximate value of $\Delta
E_T$ via $\alpha$, which leads to a simple adaptive and
computationally cheap algorithm:

If $\Delta E_T$ was very small or even negative (after changing the
optimised state by expansion of its right basis) during the current
update, we can increase $\alpha$ during the next update step on the
next site. If, on the other hand, $|\Delta E_T| \approx |\Delta E_O|$,
that is, if the error incurred during truncation nullified the gain in
energy during the optimisation step, we should reduce the value of
$\alpha$ at the next iteration to avoid making this mistake again.

In practice, it seems that keeping $\Delta E_T \approx -0.3 \Delta
E_O$ gives the fastest convergence. Given the order-of-magnitude
nature of $\alpha$, it is furthermore best to increase/decrease it via
multiplication with some factor greater/smaller than $1$ as opposed to
adding or subtracting fixed values.

Some special cases for very small $\Delta E_O$ (stuck in a local
minimum or converged to the ground state?) and $\Delta E_T > 0$ or
$\Delta E_T < \Delta E_O$ have to be considered, mostly depending on
the exact implementation.

It is unclear whether there is a causal relation between the optimal
choice of $\alpha$ and the ratio of $\Delta E_T / \Delta E_O$ or
whether both simply correlate with a proceeding DMRG calculation: at
the beginning, gains in energy are large and $\alpha$ is optimally
chosen large, whereas later on, energy decreases more slowly and
smaller values of $\alpha$ are more appropriate.

It is important to note that this is a tool to reach convergence more
quickly. If one is primarily interested in a wavefunction representing
the ground state, the calculation of a new $\alpha$ at each iteration
comes at essentially zero cost. If, however, the aim is to extrapolate
in the truncation error during the calculation, then a fixed value for
$\alpha$ is of course absolutely necessary.

\section{\label{sec:numexps}Numerical Examples}

\subsection{\label{sec:numexps:stuckspins}DMRG Stuck in a Local Minimum}
\begin{figure}
  \includegraphics[width=\columnwidth]{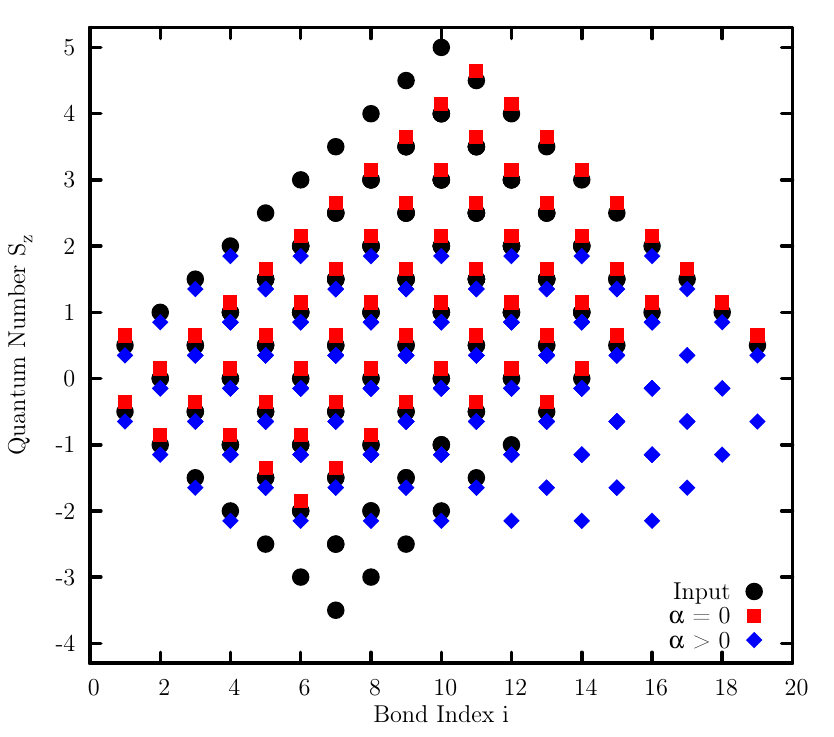}
  \caption{\label{fig:qnums-dist}(Colour online) The quantum number
    distribution as counted from the right at each bond of a $l = 20$
    system with $S=\frac{1}{2}$ and $S_z^{\mathrm{total}} = 0$. The
    artificial input state is shown with black circles. Two DMRG
    calculations have then been done on this input state, once with no
    enrichment term ($\alpha = 0$, red squares) and once with subspace
    expansion enabled ($\alpha \neq 0$, blue diamonds). It is clearly
    visible that without enrichment, DMRG3S can reduce some weights to
    zero, but cannot add new states -- red only occurs together with
    black. As soon as enrichment is enabled, DMRG3S restores $\pm S_z$
    symmetry and reflective symmetry over the 10$^\mathrm{th}$ bond
    and finds a much better ground state.}
\end{figure}

In this sub-section, we will give a short example of how DMRG can get
stuck in a local minimum even on a very small system. Consider $20$
$S=\frac{1}{2}$ spins with isotropic antiferromagnetic interactions
and open boundary conditions. The $U(1)$ symmetry of the system is
exploited on the MPS basis, with the overall $S_z$ forced to be
zero. The initial state is constructed from 20 linearly independent
states, all with $3$ sites on the very right at $S_z = 0.5$ and $m =
20$ in total. The quantum number distribution at each bond is plotted
in Fig.~\ref{fig:qnums-dist} as black circles.

DMRG3S is run with subspace expansion disabled, i.e.\ $\alpha = 0$
throughout the calculation. The algorithm ``converges'' to some
high-energy state at $E^{\alpha = 0} = -6.35479$. The resulting
quantum number distribution (red squares in Fig.~\ref{fig:qnums-dist})
shows clear asymmetry both between the left and right parts of the
system and the $+S_z$ and $-S_z$ sectors at any given bond. It is also
visible that while some states are removed by DMRG3S without enrichment,
it cannot add new states: the red squares only occur together with the
black filled circles from the input state.

If we enable enrichment via subspace expansion, i.e.\ take $\alpha
\neq 0$, DMRG3S quickly converges to a much better ground state at
$E^{\alpha \neq 0} = -8.6824724$. The quantum numbers are now evenly
distributed between the left- and right parts of the system and $\pm
S_z$ symmetry is also restored.

\subsection{\label{sec:numexps:systems}Application to Physical Systems}
In the following subsections, we will compare the two single-site DMRG
algorithms CWF and DMRG3S when applied to four different physical
systems: a $S=1$ Heisenberg spin chain with periodic boundary
conditions, a bosonic system with an optical lattice potential, a
Fermi-Hubbard model at $U=1$ and quarter-filling and a system of free
fermions at half-filling.

Each algorithm is run at three different values of $m = m_{max},
m_{max}/2, m_{max}/4$ from the same initial state and run to
convergence. This way, it is possible to both observe the behaviour of
the methods at low and high accuracies.

The usual setup in DMRG calculations of starting at small $m$ and
increasing $m$ slowly while the calculation progresses makes it
unfortunately very difficult to compare between the three
methods. This is because different methods require different
configurations to converge optimally. We therefore restrict ourselves
to fixed $m$ throughout an entire calculation, even though all methods
could be sped up further by increasing $m$ slowly during the
calculation.

Errors in energy compared to a numerically exact reference value $E_0$
are plotted as a function of sweeps and CPU time. It should be
stressed that this error in energy is not directly comparable to the
truncation error traditionally used in two-Site DMRG or the variance
$\langle \hat H^2 \rangle - \langle \hat H \rangle^2$ sometimes
considered in single-site DMRG. Even small differences in energy can
lead to vastly different physical states and reaching maximal accuracy
in energy is crucial to ensure that the true ground state has been
reached.

Furthermore, a traditional two-site DMRG (2DMRG) calculation without
perturbations is done and its error in energy and runtime to
convergence is compared to the two single-site algorithms. Here,
\emph{convergence} is defined as a normalised change in energy less
than $10^{-9}$ (for $m = m_{max}$) resp. $10^{-8}$ (for $m <
m_{max}$). The \emph{runtime to convergence} is the CPU time used
until that energy was output by the eigensolver for the first time.

All calculations were performed on a single core of a Xeon E5-2650. 

\subsubsection{\label{sec:numexps:spinchain} $S=1$ Heisenberg Chain}

Firstly, we consider a $S=1$ Heisenberg spin chain with $l=100$ sites
and periodic boundary conditions implemented on the level of the
Hamiltonian as a simple link between the first and last site:

\begin{equation}
  \hat H = \sum_{i=1}^{100} \hat S_{i} \cdot \hat S_{(i+1)\%100} \;.\label{eq:numexps:spinchain}
\end{equation}

$U(1)$ symmetries are exploited and the calculations are forced in the
$S_z = 0$ sector.

\begin{figure}
  \includegraphics[width=\columnwidth]{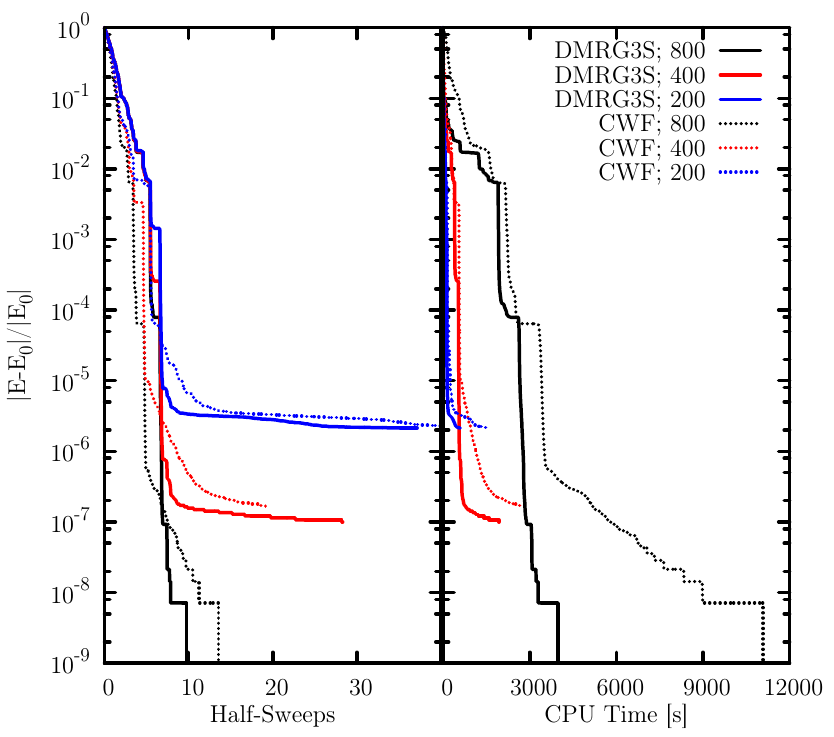}
  \caption{\label{fig:numexps:spinchain}(Colour online) Spin Chain
    Eq.~\eqref{eq:numexps:spinchain}: Normalised error in energy as a
    function of sweeps (left) and CPU time used (right) of the two
    single-site algorithms at different $m = 200, 400, 800$. DMRG3S
    shows both a speed-up and an improved convergence per sweep
    compared to CWF, with a long tail of slow convergence very visible
    for CWF at high accuracies.}
\end{figure}
\begin{table}
  \caption{\label{tab:numexps:spinchain}Spin Chain Eq.~\eqref{eq:numexps:spinchain}: Normalised error in energy at convergence and runtime to convergence of all three methods. DMRG3S is consistently faster than CWF, whereas the energies provided by 2DMRG are not comparable in accuracy.}
  \begin{tabular}{r|c|c|c}
                         & $m = 200$ & $m = 400$  & $m = 800$ \\ \hline\hline
    DMRG3S Energy Error  & $2.1\e{-6}$  & $1.0\e{-7}$   & $7.1\e{-9}$ \\
    CWF Energy Error     & $2.8\e{-6}$  & $1.7\e{-7}$   & $7.1\e{-9}$ \\
    2DMRG Energy Error   & $1.1\e{-5}$  & $8.6\e{-7}$   & $1.0\e{-7}$ \\ \hline
    DMRG3S Runtime       & \unit{583}{s} & \unit{1935}{s} & \unit{3990}{s} \\
    CWF Runtime          & \unit{1519}{s} & \unit{2695}{s} & \unit{11133}{s} \\
    2DMRG Runtime        & \unit{762}{s} & \unit{3181}{s} & \unit{21963}{s} \\\hline \hline
  \end{tabular}
\end{table}

This system is of particular interest as, firstly, it is one of the
standard benchmarking systems with well-known analytic values for the
ground-state energy. Secondly, it is a one-dimensional system where
the case of periodic boundary conditions can still be tackled by
DMRG. The larger MPO bond dimension resulting from these PBC similarly
arises during the simulation of quasi two-dimensional systems as
cylinders. The same applies to the non-nearest-neighbour interactions
in this system (between the first and last site) and cylindrical
systems.

Fig.~\ref{fig:numexps:spinchain} compares the error in energy with
respect to the reference value $E_0 = -140.148\;404$ for DMRG3S and
CWF for $m = 200, 400, 800$ as a function of sweeps and computation
time.

During the first three to four sweeps, DMRG3S exhibits a smaller
convergence rate per sweep, however, compared to the first sweeps of
CWF, they also cost negligible CPU time. Afterwards, DMRG3S offers
comparable (at medium accuracies) or much improved (at high
accuracies) convergence rate per sweep as compared to CWF together
with a still reduced average runtime per sweep. Combined, these
effects lead to a speed-up of $2.6, 1.3$ and $2.7$ for $m = 200, 400$
and $800$ respectively between CWF and DMRG3S when considering the
runtime to convergence.

In comparison, the 2DMRG algorithm does not handle the periodic
boundary conditions well and yields energies higher than the
single-site algorithms with perturbations
(cf.~Tab.~\ref{tab:numexps:spinchain}). Runtime to convergence is
hence not comparable.

\subsubsection{\label{sec:numexps:bosons}Dilute Bosons on an Optical Lattice}

We carry on to study bosons in a modulated potential of $10$ unit cells,
each with $16$ sites. The cutoff for local occupation numbers is
$n_{\mathrm{max}} = 5$, resulting in a local site dimension of $d =
6$. The Hamiltonian is given as
\begin{align}
  \hat H = & + \sum_{i=1}^{160} \hat n_i \left\{ \cos^2\left( 2 \pi
  \frac{i - 0.5}{16} \right) + \left(\hat n_i - 1\right) \right\} \nonumber \\
  & - \sum_{i=1}^{159} \left\{ \hat c^\dagger_i \hat c_{i+1}
  + \textrm{h.c.} \right\} \;.\label{eq:numexps:bosons}
\end{align}
This system should be fairly easy for DMRG to handle, as there are
only nearest-neighbour interactions. However, the large-scale order
due to the modulated potential and a very small energy penalty paid
for an uneven distribution of bosons was observed to cause badly
converged results.\cite{dolfi12:_multig_algor_tensor_networ_states}
Manual checks of the states returned by each method were hence done to
ensure a proper, equal distribution of bosons throughout the whole
system.

The state is initialised with $n = 80$ bosons in total. We allow $m =
50, 100, 200$ states and use the energy reference value $E_0 =
-103.646\;757$. All algorithms converge to this value at $m = 200$.

\begin{figure}
  \includegraphics[width=\columnwidth]{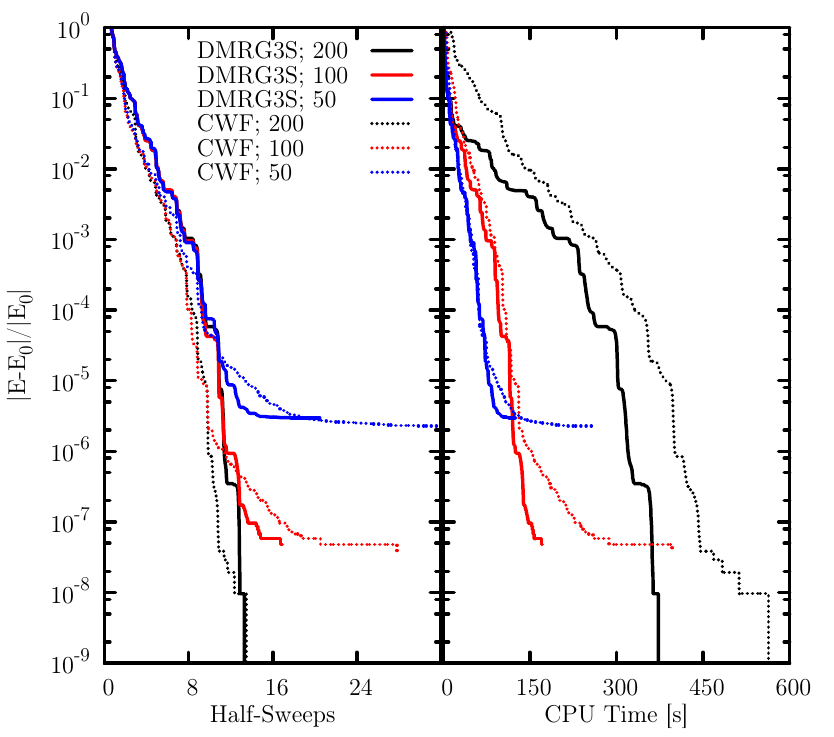}
  \caption{\label{fig:numexps:bosons}(Colour online) Bosonic System Eq.~\eqref{eq:numexps:bosons}: Normalised error in energy from
    CWF and DMRG3S as a function of sweeps (left) and CPU time used
    (right) for $m = 50, 100, 200$. Again, an improved convergence
    behaviour at high accuracies can be observed, in particular at
    smaller values of $m$. The small bond dimensions lead to a smaller
    speed-up due to faster numerical operations, which only becomes
    visible at $m = 200$.}
\end{figure}

Fig.~\ref{fig:numexps:bosons} compares CWF and DMRG3S whereas
Tab.~\ref{tab:numexps:bosons} additionally lists 2DMRG. Since the bond
dimensions are relatively small, we do not expect a speed-up from
faster numerical operations. Instead, the improved convergence
behaviour per sweep is responsible for the speed-up of $2$ of DMRG3S
over CWF at small $m$. At larger $m$, CWF converges better, but
numerical operations also become cheaper for DMRG3S for a speed-up of
$2$ again.

As there are no long-range interactions, 2DMRG also fares well with
regard to energy accuracy. However, it takes longer to converge than
the single-site methods especially at large $m$, mainly because the
eigenvalue problem in two-site DMRG is of dimension $d$ larger than in
single-site DMRG. A comparison between DMRG3S and 2DMRG leads to a
speed-up of up to $3.3$ for the case of $m = 200$.

\begin{table}
  \caption{\label{tab:numexps:bosons}Bosonic System Eq.~\eqref{eq:numexps:bosons}: Normalised error in energy at convergence and runtime to convergence of all three methods. DMRG3S is again the fastest method with a very constant speed-up of $2$ over CWF and up to $3.3$ over 2DMRG.}
  \begin{tabular}{r|c|c|c}
                         & $m = 50$ & $m = 100$  & $m = 200$ \\ \hline\hline
    DMRG3S Energy Error  & $2.9\e{-6}$  & $4.8\e{-8}$   & $<10^{-9}$ \\
    CWF Energy Error     & $2.3\e{-6}$  & $3.9\e{-8}$   & $<10^{-9}$ \\
    2DMRG Energy Error   & $1.9\e{-6}$  & $2.8\e{-8}$   & $<10^{-9}$ \\ \hline
    DMRG3S Runtime       & \unit{124}{s} & \unit{171}{s} & \unit{469}{s} \\
    CWF Runtime          & \unit{260}{s} & \unit{397}{s} & \unit{951}{s} \\
    2DMRG Runtime        & \unit{210}{s} & \unit{462}{s} & \unit{1550}{s} \\\hline \hline
  \end{tabular}
\end{table}

\subsubsection{\label{sec:numexps:fermi-hubbard}
  Fermi-Hubbard Model}

As a third example, substantially more expensive calculations are
carried out for a substantially stronger entangled Fermi-Hubbard model
of 100 sites with Hamiltonian
\begin{equation}
  \hat H = \sum_{i=1}^{100} \left\{ -\sum_{\sigma=\uparrow,\downarrow}
  \left[ \hat c^\dagger_{i,\sigma} \hat c_{i+1,\sigma} + \textrm{h.c.}\right]
  + \hat n_{i,\uparrow} \hat n_{i,\downarrow}\right\} \;. \label{eq:numexps:fermi-hubbard}
\end{equation}

Both $U(1)_{\textrm{charge}}$ and $U(1)_{\mathrm{Sz}}$ symmetries
are employed, with $50$ fermions and $S_z^\mathrm{total} = 0$ enforced
through the choice of initial state. Together with the free fermions
from the next section, we can use this system to study how criticality
and increased entanglement affect the three methods.

Calculations are done for $m = 300, 600, 1200$. All methods converge
to the same value $E_0 = -84.255\;525\;4$ at large $m$.

\begin{figure}
  \includegraphics[width=\columnwidth]{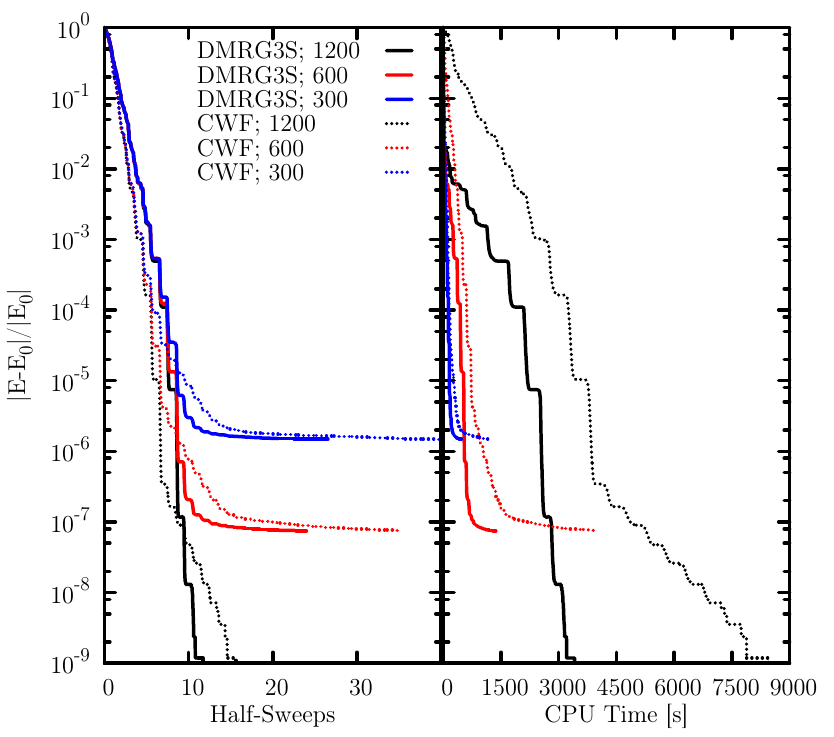}
  \caption{\label{fig:numexps:fermi-hubbard}(Colour online)
    Fermi-Hubbard Eq.~\eqref{eq:numexps:fermi-hubbard}: Normalised
    error in energy from DMRG3S and CWF as a function of sweeps
    (left) and CPU time used (right) for different bond dimensions $m
    = 300, 600, 1200$. The same basic behaviour as for the previous
    systems is repeated, with both improved convergence behaviour at
    high accuracies and faster numerical operations.}
\end{figure}
\begin{table}
  \caption{\label{tab:numexps:fermi-hubbard}Fermi-Hubbard Eq.~\eqref{eq:numexps:fermi-hubbard}: Normalised error in energy at convergence and runtime to convergence of all three methods. Accuracies are comparable between the different methods, but runtimes vary greatly.}
  \begin{tabular}{r|c|c|c}
                         & $m = 300$ & $m = 600$  & $m = 1200$ \\ \hline\hline
    DMRG3S Energy Error  & $1.5\e{-6}$  & $7.5\e{-8}$   & $<10^{-9}$ \\
    CWF Energy Error     & $1.5\e{-6}$  & $7.6\e{-8}$   & $<10^{-9}$ \\
    2DMRG Energy Error   & $1.3\e{-6}$  & $6.4\e{-8}$   & $<10^{-9}$ \\ \hline
    DMRG3S Runtime       & \unit{474}{s} & \unit{1367}{s} & \unit{3955}{s} \\
    CWF Runtime          & \unit{1215}{s} & \unit{3917}{s} & \unit{10122}{s} \\
    2DMRG Runtime        & \unit{727}{s} & \unit{2950}{s} & \unit{15596}{s} \\\hline \hline
  \end{tabular}
\end{table}

Fig.~\ref{fig:numexps:fermi-hubbard} compares the two single-site
methods while Tab.~\ref{tab:numexps:fermi-hubbard} summarises all
three DMRG implementations. Since the system only exhibits local interactions, 2DMRG fares well
and all methods generally provide comparable energies. The difference
is therefore in the runtime needed to achieve these energies. Compared
to CWF, DMRG3S achieves a speed-up of $\approx 2.6$ consistently at
all $m$, as the smallest $m = 300$ is already large enough to justify
the assumption $d^2 w \ll m$ in the speed-up of numerical operations.
In particular, it continues to converge quickly at high accuracies
whereas CWF develops a long tail of slow convergence. The speed-up
compared to 2DMRG is smaller at lower values of $m$, but increases to
$3.9$ at $m = 1200$.

\subsubsection{\label{sec:numexps:fermions}Free Fermions}

\begin{figure}
  \includegraphics[width=\columnwidth]{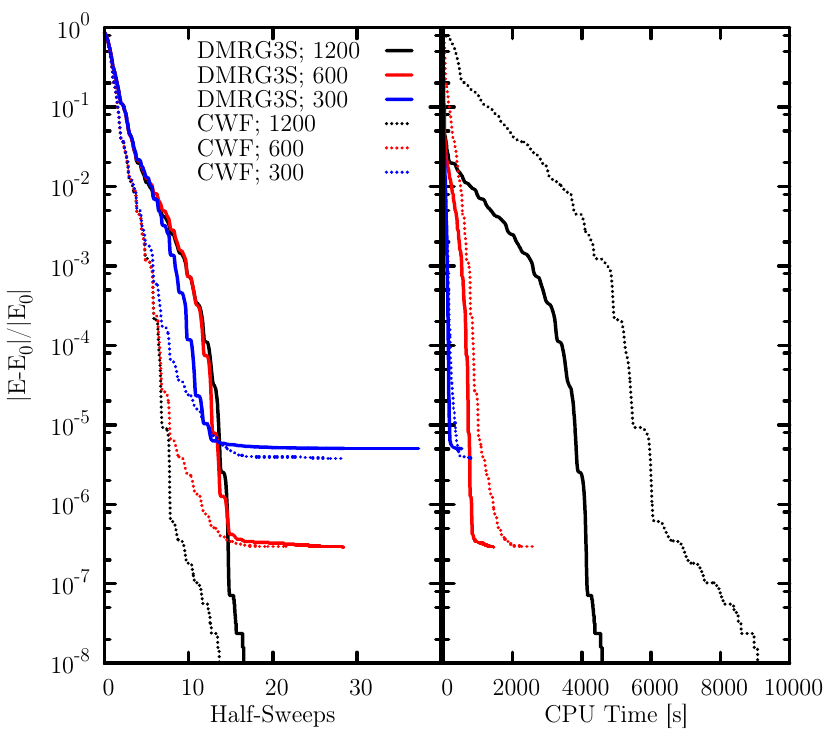}
  \caption{\label{fig:numexps:fermions}(Colour online) Free Fermions
    Eq.~\eqref{eq:numexps:fermions}: Normalised error in energy from
    CWF and DMRG3S as a function of sweeps (left) and CPU time used
    (right) at $m = 300, 600, 1200$. CWF again exhibits a long tail of
    slow convergence while DMRG3S converges quickly at all $m$ and all
    accuracies.}
\end{figure}

\begin{table}
  \caption{\label{tab:numexps:fermions}Free Fermions Eq.~\ref{eq:numexps:fermions}): Normalised error in energy at convergence and runtime to convergence of all three methods.}
  \begin{tabular}{r|c|c|c}
                         & $m = 300$ & $m = 600$  & $m = 1200$ \\ \hline\hline
    DMRG3S Energy Error  & $5.0\e{-6}$  & $2.8\e{-7}$   & $<10^{-9}$ \\
    CWF Energy Error     & $3.8\e{-6}$  & $2.8\e{-7}$   & $<10^{-9}$ \\
    2DMRG Energy Error   & $3.7\e{-6}$  & $2.6\e{-7}$  & $<10^{-9}$ \\ \hline
    DMRG3S Runtime       & \unit{533}{s} & \unit{1452}{s} & \unit{4643}{s} \\
    CWF Runtime          & \unit{863}{s} & \unit{2590}{s} & \unit{9586}{s} \\
    2DMRG Runtime        & \unit{794}{s} & \unit{4584}{s} & \unit{29698}{s} \\\hline \hline
  \end{tabular}
\end{table}

Finally, we consider a model of free fermions on a chain of 100 sites
with Hamiltonian
\begin{equation}
  \hat H = - \sum_{i=1}^{100} \sum_{\sigma=\uparrow,\downarrow}
  \left[ \hat c^\dagger_{i,\sigma} \hat c_{i+1,\sigma} + \textrm{h.c.}\right]\;. \label{eq:numexps:fermions}
\end{equation}
The maximally delocalised wavefunction found in the ground-state of
this system is notoriously difficult for MPS formats in general to
reproduce faithfully. At the same time, most other parameters are
identical ($d$, $l$, $m$) or very close ($w$) to those in the
Fermi-Hubbard model from Section~\ref{sec:numexps:fermi-hubbard}. The
calculation is done using $U(1)_{\textrm{charge}}$ and
$U(1)_{\mathrm{Sz}}$ symmetries at half-filling with $N = 100$
fermions and $S_z^{\mathrm{total}} = 0$. The choice of $m$ is the same
as for the Fermi-Hubbard system, namely $m = 300, 600, 1200$. We used
$E_0 = -126.602\;376$ as the reference value, since all methods
converged to this ground-state energy at $m = 1200$.

The results in Tab.~\ref{tab:numexps:fermions} and
Fig.~\ref{fig:numexps:fermions} mostly follow the previous results for
locally interacting systems: Accuracies of all methods are essentially
identical, whereas time to convergence varies between the methods. At
small $m$, there are some speed-ups of DMRG3S over CWF, largely
due to better convergence behaviour per sweep, whereas a significant
advantage of DMRG3S becomes visible at larger $m$, when numerical
operations become cheaper compared to the CWF method. Correspondingly,
the speed-up from CWF to DMRG3S increases from $1.6$ at $m = 300$ to
$2$ at $m = 1200$.

Similarly, the larger numerical cost of two-site DMRG becomes more
noticeable at larger $m$, with the speed-up between 2DMRG and DMRG3S
increasing from $1.5$ at $m = 300$ to more than $6$ at $m = 1200$. 

Compared to the non-critical Fermi-Hubbard system from
Section~\ref{sec:numexps:fermi-hubbard}, we observe larger errors in
energy at fixed $m$, as expected. Correspondingly, as more eigenvalues
contribute significantly, convergence of both the eigenvalue solver
and the singular value decompositions becomes slower, leading to a
slow-down of all three methods.

\section{\label{sec:conclusions}Conclusions}
The new strictly single-site DMRG (DMRG3S) algorithm results in a
theoretical speed-up of $\sim (d+1)/2$ during the optimisation steps
compared to the centermatrix wavefunction formalism (CWF), provided
that $d^2 w / m$ is small. Further, convergence rates per sweep are
improved in the important and computationally most expensive
high-accuracy/large-$m$ phase of the calculation. In addition,
auxiliary calculations (enrichment, normalisation, etc.) are sped up
and memory requirements are relaxed.

Numerical experiments confirm a speed-up within the theoretical
expectations compared to the CWF method. The efficiency of single-site
DMRG in general compared to the traditional two-site DMRG was
substantiated further by a large speed-up at comparable accuracies in
energy.

\begin{acknowledgments}
  We would like to thank S.~Dolgov, D.~Savostyanov and I.~Kuprov
  for very helpful discussions. C.~Hubig acknowledges funding through the
  ExQM graduate school and the Nanosystems Initiate Munich. F.~A.~Wolf
  acknowledges support by the research unit FOR 1807 of the DFG.
\end{acknowledgments}

\end{document}